\documentclass[acs,preprint,showkeys]{revtex4-1}
\usepackage{graphicx}
\usepackage{color}
\usepackage{pdfpages}

\begin{document}

\title{Thermal effects on CH$_3$NH$_3$PbI$_3$ perovskite from {\em ab-initio} molecular dynamics simulations}

\author{Marcelo A. Carignano}
\email{mcarignano@qf.org.qa}
\affiliation{Qatar Environment and Energy Research Institute, P.O. Box 5825, Doha, Qatar}

\author{Ali Kachmar}
\affiliation{Qatar Environment and Energy Research Institute, P.O. Box 5825, Doha, Qatar}

\author{J\"urg Hutter}
\affiliation{Department of Chemistry, University of Zurich, Winterthurerstrasse 190, CH-8057, Zurich, Switzerland}

\date{September 16, 2014}

\begin{abstract}
We present a molecular dynamics simulation study of CH$_3$NH$_3$PbI$_3$ based on forces calculated from density functional theory. The simulation were performed on model systems having 8 and 27 unit cells, and for a total simulation time of 40 ps in each case. Analysis of the finite size effects, in particular the mobility of the organic component, suggests that the smaller system is over correlated through the long range electrostatic interaction. In the larger system this finite size artifact is relaxed producing a more reliable description of the anisotropic rotational behavior of the methylammonium molecules. The thermal effects on the optical properties of the system were also analyzed. The HOMO-LUMO energy gap fluctuates around its central value with a standard deviation of approximately 0.1 eV. The projected density of states consistently place the Fermi level on the $p$ orbitals of the I atoms, and the lowest virtual state on $p$ orbitals of the Pb atoms throughout the whole simulation trajectory.  
\end{abstract}

%\keywords{Perovskites}

\maketitle

\newpage

\section{Introduction}

In the recent years there has been a spike of interest on hybrid organic-inorganic perovskite materials\cite{Editorial:2014aa,McGehee:2014aa} due to an impressive series of light to electricity efficiency conversion improvements \cite{Kojima:2009aa,Im:2011aa,Chung:2012aa,Kim:2012aa,Lee:2012aa}. Indeed, in just over five years of intense research activity the efficiency of CH$_3$NH$_3$PbI$_3$ is (from now on referred to as MAPbI$_3$) above 17 \% \cite{Im:2014aa}. The experimental efforts have reached a large degree of understanding on the structural, dynamic and electronic structure properties of the material. The MAPbI$_3$ perovskite has three distinct phases below melting at $T \simeq 470$ K \cite{Cohen:2014aa}. The low temperature orthorhombic phase transforms to a tetragonal structure at 161.4 K, which is stable up to 330.4 K \cite{Stoumpos:2013aa,Baikie:2013aa}. The high temperature phase is cubic. The MA molecules have rotational degrees of freedom in both, the tetragonal and cubic phases. Therefore, these two phases are indeed plastic phases \cite{Sherwood:1979aa,Wasylishen:1985aa} with some similarities to organic ionic plastic crystals \cite{Carignano:2013aa}. In this study we concentrate in the intermediate tetragonal phase that is the stable phase at standard ambient conditions.

From a theoretical and computational point of view there has been a number of studies on MAPbI$_3$ and similar compounds addressing electronic structure properties with static density functional theory (DFT) calculations \cite{Mosconi:2013aa,Frost:2014aa,Umari:2014aa,Wang:2014aa,Even:2014aa,Even:2013aa,Filippetti:2014aa}. Spin-orbit coupling has been shown to substantially affect the energy band gap \cite{Even:2013aa}. Up to date, three studies based on {\em ab-initio} molecular dynamics simulations have been reported \cite{Lindblad:2014aa,Mosconi:2014aa,Frost:2014ab}. The first one by Lindblad et al. \cite{Lindblad:2014aa} is a simulation performed with CPMD on the cubic phase represented with 8 unit cells and addressing mainly electronic structure properties. The second work by Mosconi et al. \cite{Mosconi:2014aa} is based on a similar small system but targeting also structural and dynamic properties. The third work by Frost et al. \cite{Frost:2014ab}, is also based on a system of similar characteristics in terms of structure and size and addressing the anisotropy in the rotations of the MA molecules and the formation of microscopic polarization domains. The first two studies \cite{Lindblad:2014aa,Mosconi:2014aa} were conducted on systems of 96 atoms initialized in the cubic structure and simulated for 6 ps and 13 ps, respectively. In fact, in Ref. \cite{Mosconi:2014aa} the positions of the Pb atoms are frozen in one case, and the positions of Pb and I atoms are frozen in a second simulation case. In this way the authors investigate the mobility of the MA within the inorganic cage. The third work \cite{Frost:2014ab} was conducted on a cubic structure with 80 atoms with no constraints in the positions of the inorganic atoms and for much longer times, using 58 ps for the analysis. Even though these works represent important contributions, it is clear that simulation studies on larger system having the proper crystal structure at the target temperature and for sufficiently long times must be undertaken in order to gain an unbiased understanding of the coupling between the electronic and nuclear degrees of freedom for this class of systems.

\begin{figure}[!t]
\includegraphics*[width=0.35\textwidth]{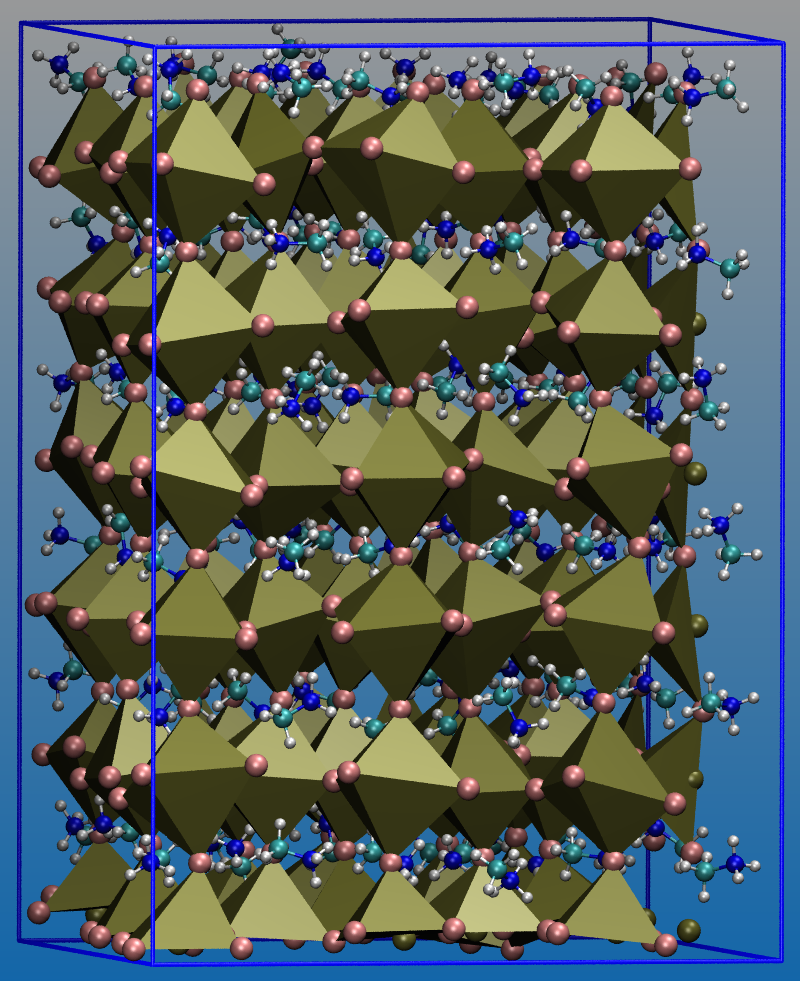}
\caption{Snapshot of System 333 at $t=$40 ps. The I atoms are represented in pink, N in blue, C in cyan and H in white. The Pb atoms are not visible, hidden by the polyhedra representation.}
\label{view}
\end{figure}

Here we present an {\em ab-initio} molecular dynamics simulation study on the tetragonal phase of MAPbI$_3$ without imposing any constraint in the position of the atoms. Our findings show that the simulated system has to be sufficiently large in order to avoid excessive finite size effects, and that the simulated times needs to be long enough to relax the initial structure to one that properly corresponds to the simulated temperature. The expensive nature of this type of simulations prohibit the exploration of large time scales. Nevertheless, we clearly show that the first 5 ps of the trajectories are subject to excessive fluctuations and must be discarded for the analysis. Moreover, the high mobility of the MA molecules is strongly affected by the size of the system. The rest of paper is organized in three sections describing the computational methodology, the presentation and discussion of the results of the simulations and the final conclusions. A Supplementary Material is also provided with more details and analysis.

\section{Models and Computational Methodology}

We have considered two different system sizes: A smaller system with 8 unit cells (384 atoms) and a larger system with 27 unit cells (1296 atoms). Note that the unit cell corresponding to the tetragonal phase contains 48 atoms. We refer to the smaller system as System 222 because it was prepared by replicating the experimental unit cell twice on each Cartesian direction. Analogously, we refer to the larger system as System 333. A snapshot of the larger system is shown in Figure \ref{view}.
The initial coordinates for the preparation of the systems were taken from the experimental data  obtained a 293 K and recently reported by Stoumpos et al. \cite{Stoumpos:2013aa}. The crystal structure corresponds to the I4cm symmetry space group, with unit cell dimensions $a=b=8.849(2)$ \AA\ and $c=12.642(2)$ \AA. 

The simulations have been carried out using the CP2K package \cite{VandeVondele:2005aa,Hutter:2014aa}. We have used the hybrid Gaussian and plane waves method (GPW) as implemented in the Quickstep module of the CP2K package. The control of the temperature has been implemented for both the ionic and electronic degrees of freedom by using Nos\'e-Hoover thermostats \cite{Nose:1984aa,Hoover:1986aa} with 3 chains with a target temperature of 293 K and a time constant of 50 fs. The pressure was maintained using an isotropic algorithm with a reference pressure of 1 atm \cite{Martyna:1996aa,Mundy:2000aa}. The electronic structure properties were calculated using the PBE functional with the Grimme correction scheme to account for the dispersion interactions \cite{Grimme:2006aa,Grimme:2010aa}. Kohn-Sham orbitals are expanded in a Gaussian basis set (DZVP-MOLOPT for Pb, I, C, N, H) \cite{VandeVondele:2007aa} and we employed the norm-conserving GTH pseudopotentials \cite{Goedecker:1996aa,Krack:2005aa}. The auxiliary plane wave (PW) basis set was defined by the energy cutoff of 300 Ry. The time step of the integration of the dynamic equations was set to 1 fs. The simulations were extended up to 40 ps for both systems.

\section{Results}

\begin{figure}[!b]
\includegraphics*[scale=0.6]{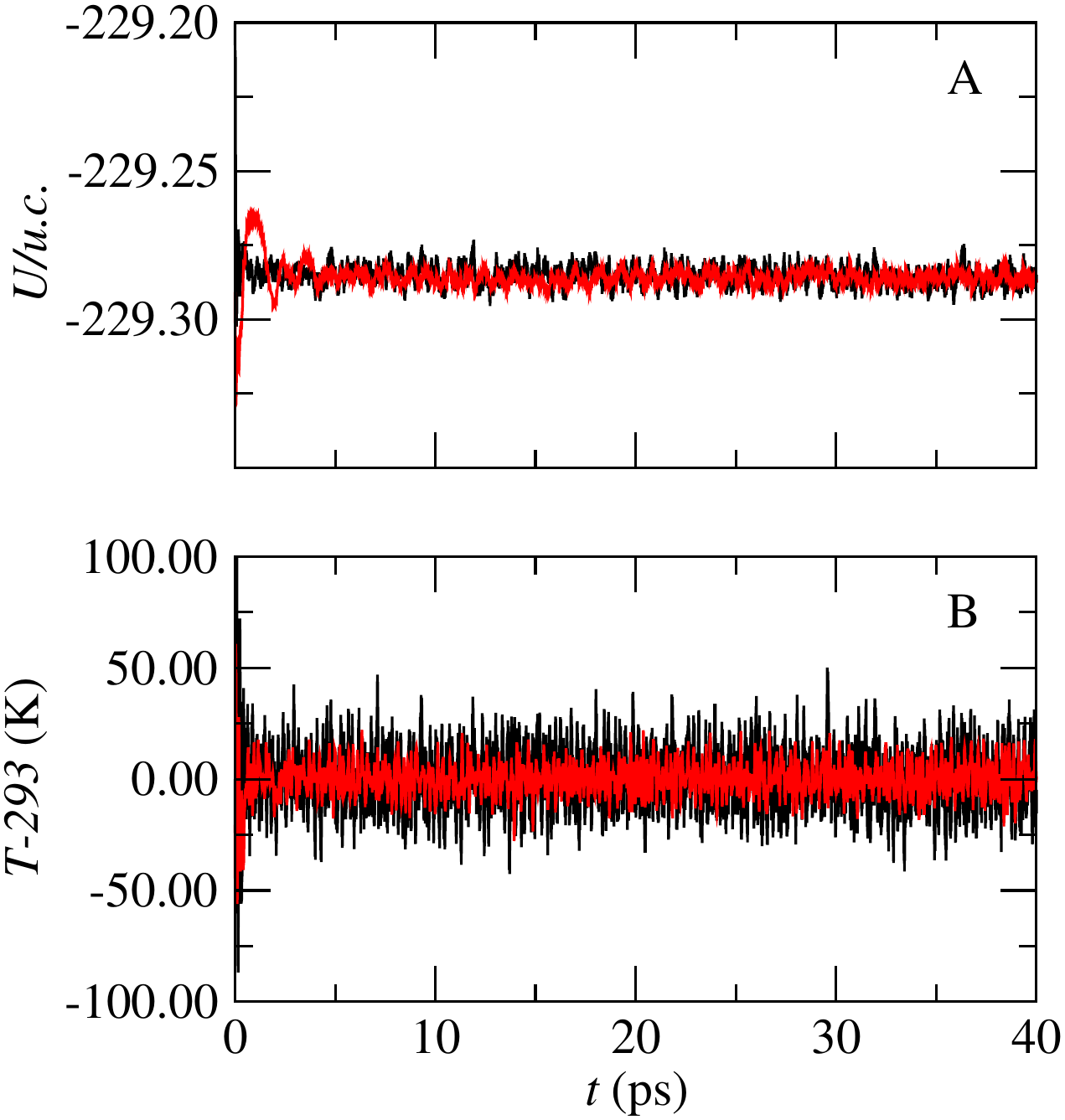}
\caption{A) Potential energy per unit cell as a function of time for the two simulated systems. Due to the use of pseudo potentials the energy units are not absolute and therefore have been dropped. B) Temperature of the system relative to the target thermostat temperature as a function of time. In both panels, the black lines correspond to System 222 and the red line to System 333.}
\label{ut}
\end{figure}

We start the analysis of the simulations by looking at the stability of the potential energy and the temperature for the two simulated systems. This is shown in Figure \ref{ut}, which presents the potential energy divided by the number of unit cells in each simulation box in the top panel and the temperature in the bottom panel. The plots represent the results from the beginning of the simulations and therefore the equilibration/thermalization stage is also included in the represented 40 ps. For most of the trajectory both systems show the same stable behavior. Nevertheless, the initial period is characterized by relatively large fluctuations. Indeed, System 333 clearly shows strong oscillations, especially in the potential energy, that relaxes at about 5 ps to its average limiting value. The oscillating pattern is also observed in System 222, although it expands over a shorter period of time. The instantaneous  temperature of the systems (or the kinetic energy) also reflects this transient initial stage through fluctuations around the target thermostat value that are larger at the initial steps than after 5 ps. The fluctuations at the beginning of the simulations are not surprising and their origin can be traced back to the unrealistic initial low entropy of the systems that needs to increase in order to reach its thermodynamic value. Namely, the initial conformation has all the methylammonium groups oriented in the  same way (along the $z$ direction) and this is not an adequate representation of the system at ambient temperature. Therefore, the re-accommodation of the methylammonium molecules occurs affecting the size of the simulation box (see below) and the instantaneous values of the kinetic and potential energies. 

\begin{figure}[!t]
\includegraphics*[scale=0.6]{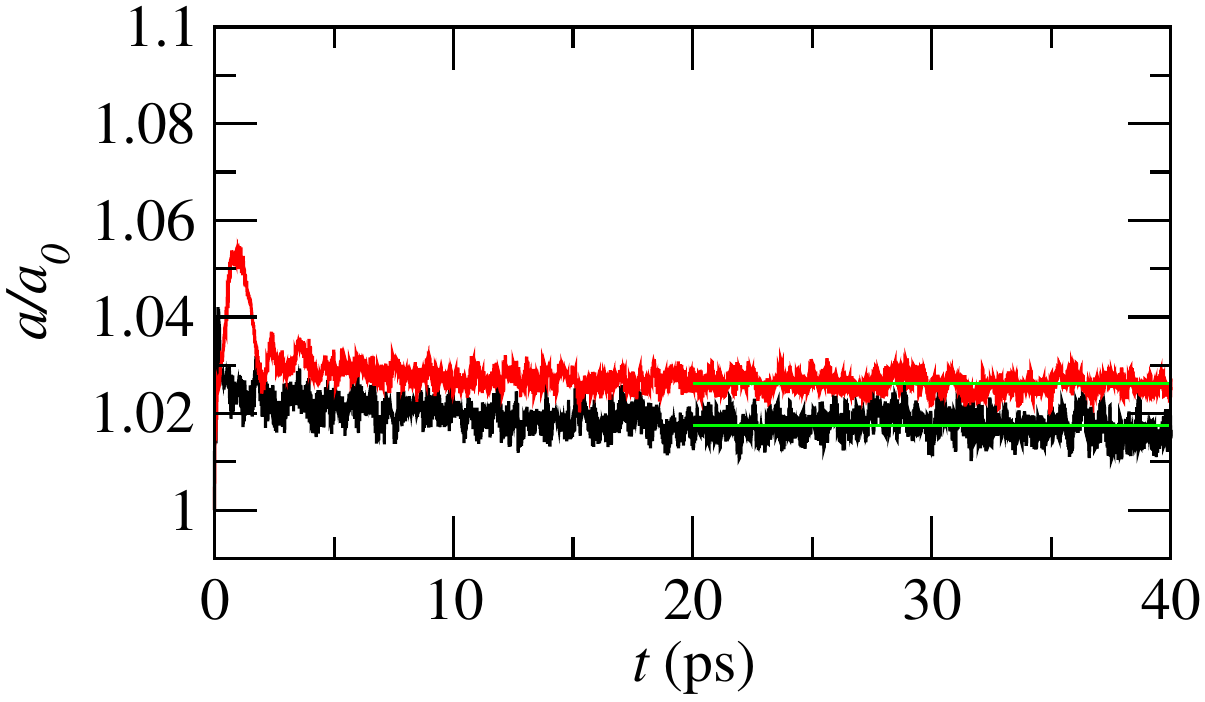}
\caption{Time evolution of the relative lattice parameter for System 222 (black line) and System 333 (red line). The horizontal lines represent the average values in the second half of the simulations, which are 1.017 and 1.026.}
\label{relative-box}
\end{figure}

Next we analyze the overall structural stability by looking at the lattice parameters as a function of time, shown in Figure \ref{relative-box}. Since the simulations were run imposing a constraint on the relative box dimensions it is sufficient to look at only one of the relative lattice parameters. (The validity of this approach was tested by freeing the isotropic constraint and no significant effects were observed, see Supplemental Information.) The initial large fluctuations displayed by the instantaneous values of the kinetic and potential energies are reflected here by an initial expansion of the simulation box. System 333 displays stronger initial fluctuations than System 222 and both systems show a slow relaxation towards what {\em appears} to be an equilibrated state. 
Although both systems expand with respect to their initial dimension, the larger system expands slightly more than the smaller one. The difference in the limiting values of the lattice parameter may just be a reflection of the importance of the entropic contribution of the MA molecules, which may require a system size larger than System 222 as it will be shown below in connection with the relaxation times for the MA rotations.

\begin{figure}[!t]
\includegraphics*[scale=0.6]{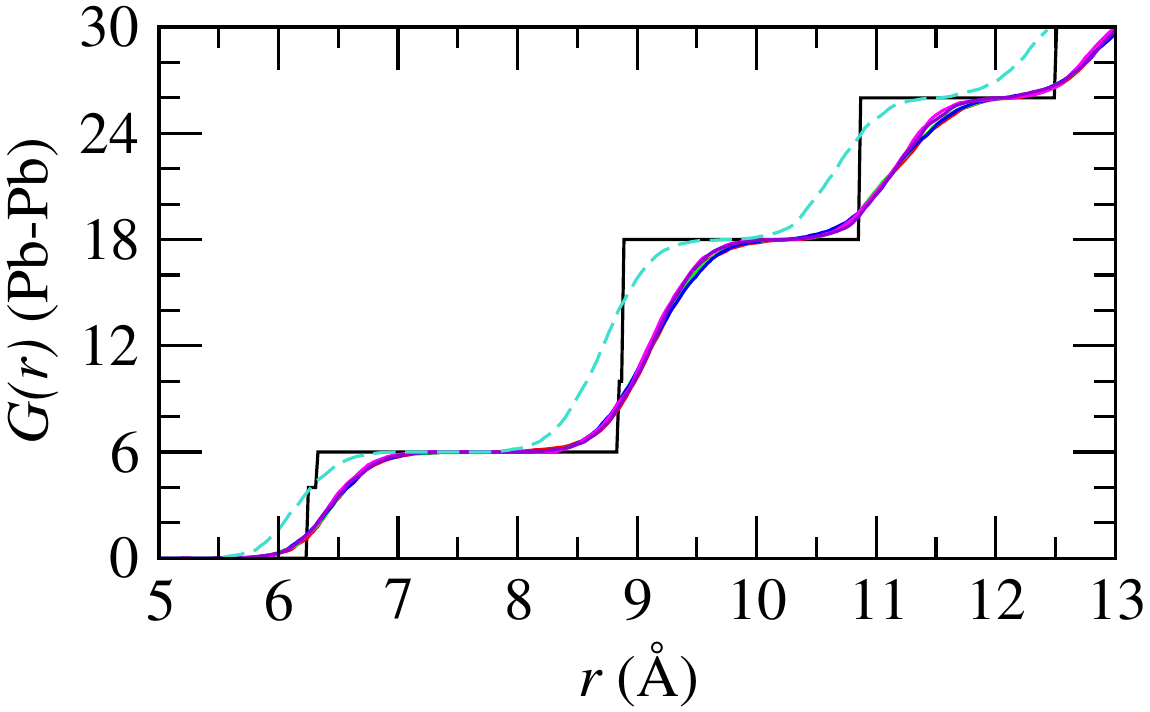}
\caption{Cumulative pair distribution function between Pb atoms for System 333. The black line shows the characteristic stepwise $G(r)$ of the perfect crystal. There are five more lines that are practically identical and correspond to different times during the simulations as follow: 5 ps (red), 10 ps (green), 20 ps (blue) 30 ps (magenta) and 40 ps (violet). The turquoise dashed line represents the results at 40 ps with $r$ scaled by 1.026, which is the lattice expansion factor relative to the initial perfect crystal.}
\label{GrPbPb}
\end{figure}

We continue the analysis by looking at the local structure and its stability throughout the simulation. In particular, we focus on the short range coordination of the different atoms as displayed by the cumulative radial distribution function $G(r)=\int 4 \pi r^2 g(r) dr$. $G(r)$ directly shows the number of neighbors of the central atom, and therefore it is an efficient way to quantitatively account for local structural details. The small size of System 222 allows for the proper computation of pair distribution functions up to $r \sim 8.9$ \AA, which is just enough to see the first neighbors and the onset of the second neighbors. Here we display the predictions of System 333, which allows for the proper consideration of $r$ values up to $\sim$13.2 \AA. In Figure \ref{GrPbPb} we show the cumulative pair distribution function between Pb atoms. For reference, we plot the corresponding function for the perfect crystal that is a stepwise curve. The $G(r)$ calculated from the simulations were obtained averaging 100 fs (100 frames) around different times along the simulation trajectory, from 5 ps to 40 ps. All the curves coalesce into a single pattern that smoothly follows the prefect crystal result very well. The temperature effects are reflected by the smooth transitions between different sequence of neighbors as the distance increases. In the perfect crystal, each Pb atom has six first neighbors and 12 second neighbors forming a (quasi) simple cubic structure with first neighbors at 6.3 \AA\ and second neighbors at $\sqrt{2}~6.3 \simeq 8.9$ \AA. The corresponding distances for the simulation results are 2.6 \% larger due to the expanded lattice parameters. Figure \ref{GrPbPb} also shows the results at 40 ps with the distance scaled by $1/1.026$ in order to account for the expansion of the simulation box relative to the initial perfect system. 

\begin{figure}[!t]
\includegraphics*[scale=0.6]{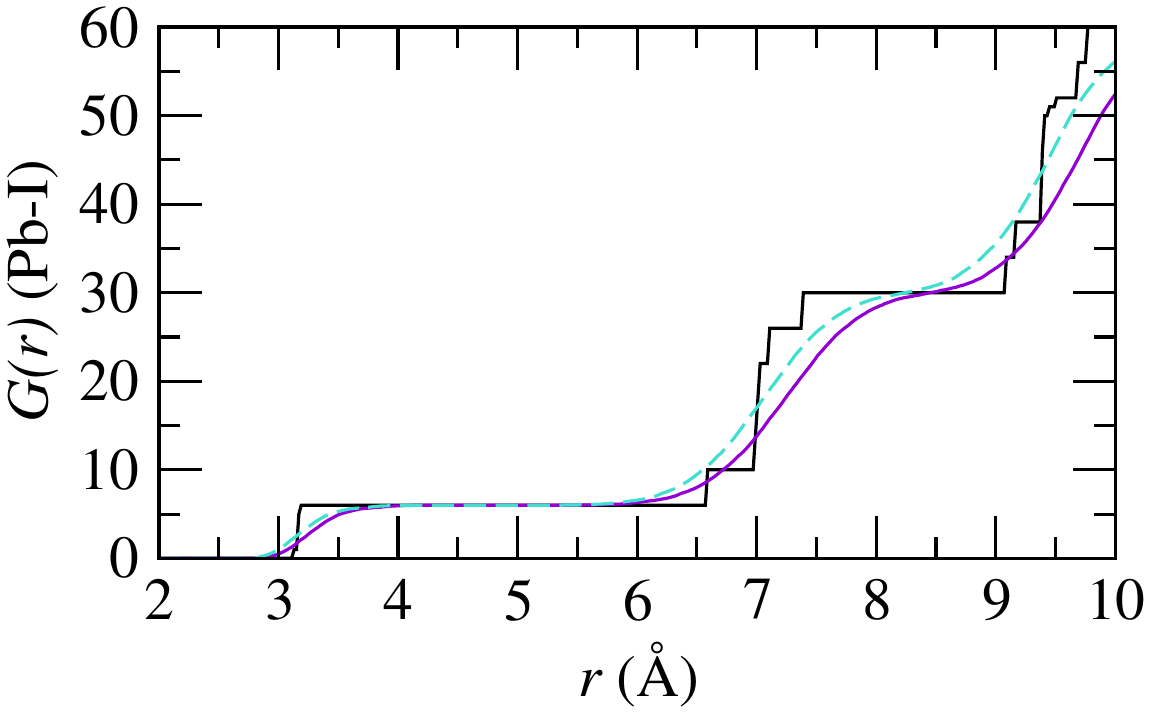}
\caption{Cumulative pair distribution function for Pb-I atoms calculated on System 333. The violet line represents the results at 40 ps, the dashed turquoise line is the same curve with the distances scaled by a factor 1.026 and the black line represents the perfect crystal structure.}
\label{GrPbI}
\end{figure}

The $G(r)$ for the Pb-I pairs is displayed in Figure \ref{GrPbI}. Again in this case the simulations results  smoothly follow the ideal crystalline curve. Each Pb atom has six I atoms first neighbors, resembling the conformation of a binary {\em fcc} structure where the Pb are at the corners and the I atoms are at the faces' centers. A layer of 24 second neighbors emerges at distances corresponding to the smaller size of the unit cell. The results of both distribution functions centered at the Pb atoms clearly show that the overall crystalline structure of the inorganic components of the perovskite is stable throughout the simulations and therefore we are confident that the employed DFT scheme and dynamic approaches are adequate to describe the thermodynamic and dynamic properties of the system.

\begin{figure}[!b]
\includegraphics*[scale=0.6]{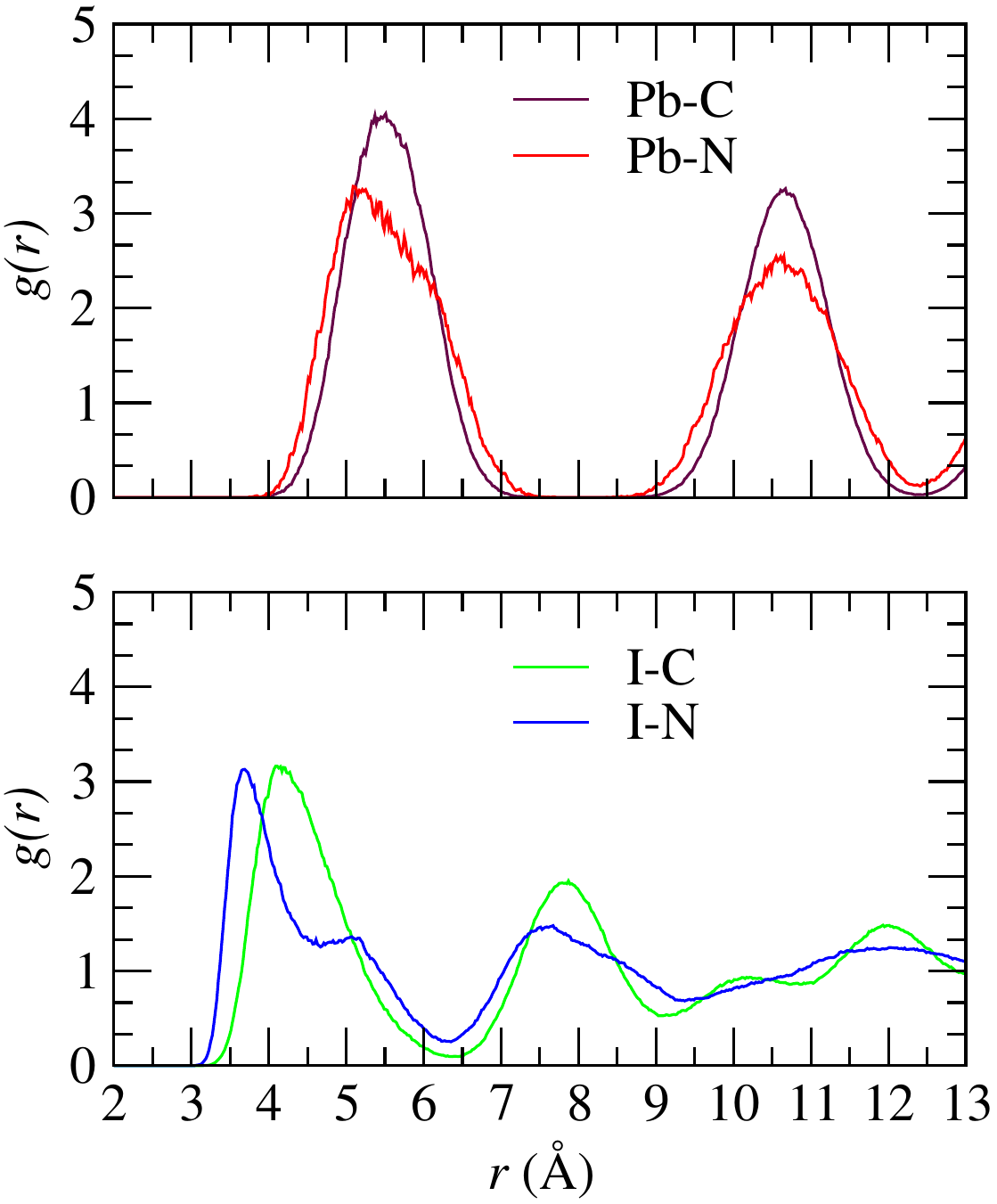}
\caption{Pair distribution function for Pb-C and Pb-N (top panel), and I-C and I-N (bottom panel) atom pairs as indicated in the figure.}
\label{GrPbCN}
\end{figure}

\begin{figure}[!t]
\includegraphics*[scale=0.6]{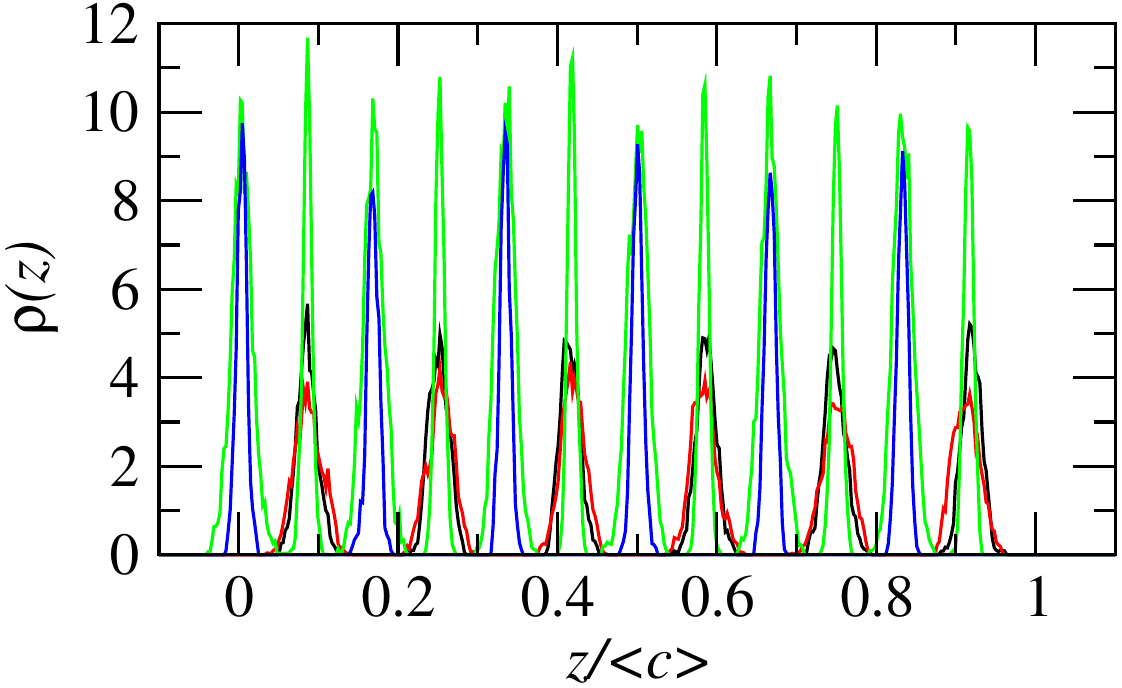}
\caption{Density profiles along the $c$ direction for C (black), N (red), I (green) and Pb (blue). The density scale is arbitrary, it only preserves the relative population of the different species.}
\label{profile}
\end{figure}

The methylammonium molecules have the possibility to rotate in their confining cages defined by the Pb and I atoms of the nearly stable binary {\em fcc} structure. As a consequence of the rotation, the distance from the N and C atoms to the Pb and I atoms can vary within the confinement range. For that reason, it is useful in this case to look at the pair distribution functions $g(r)$ displayed in Figure \ref{GrPbCN}. The mobility of the MA molecule translates into broad peaks in the pair distribution functions. The curves corresponding to the N atoms show a less symmetric first peak than the curves corresponding to the C atoms. The sharp peak for the I-N pair distribution function is the result of hydrogen bonding. The analysis of the hydrogen bonds using the standard geometrical criteria with a cutoff of 4.6 \AA for the I-N distance and 20$^\circ$ for the I-N-H angle reveals an average of 53.6 bonds; i.e. almost half of the MA molecules are participating in hydrogen bonds at any given time (see Supplementary Information). It is also interesting to note that the formation of an hydrogen bond requires that the N atoms are close to the faces' centers, a location that is not favorable from steric considerations since the large spaces are along the cage diagonals.
The different role played by the C and N atoms is also reflected, although slightly, by the density profiles showed in Figure \ref{profile}. The densities were calculated over the last 5 ps of simulation and are normalized so that the integral along the $z$ direction results in 1 for Pb, C and N, and 3 for I. The layering of the material is clearly reflected by the succession of peaks, with the I atoms showing their placement in the same planes as the Pb atoms and also at the intermediate planes. The C and the N atoms are only at the intermediate planes between the Pb atoms, and the peaks corresponding to the N atoms are slightly broader than the peaks corresponding to the C atoms, reflecting the hydrogen bond formation.

\begin{figure}[!t]
\includegraphics*[scale=0.6]{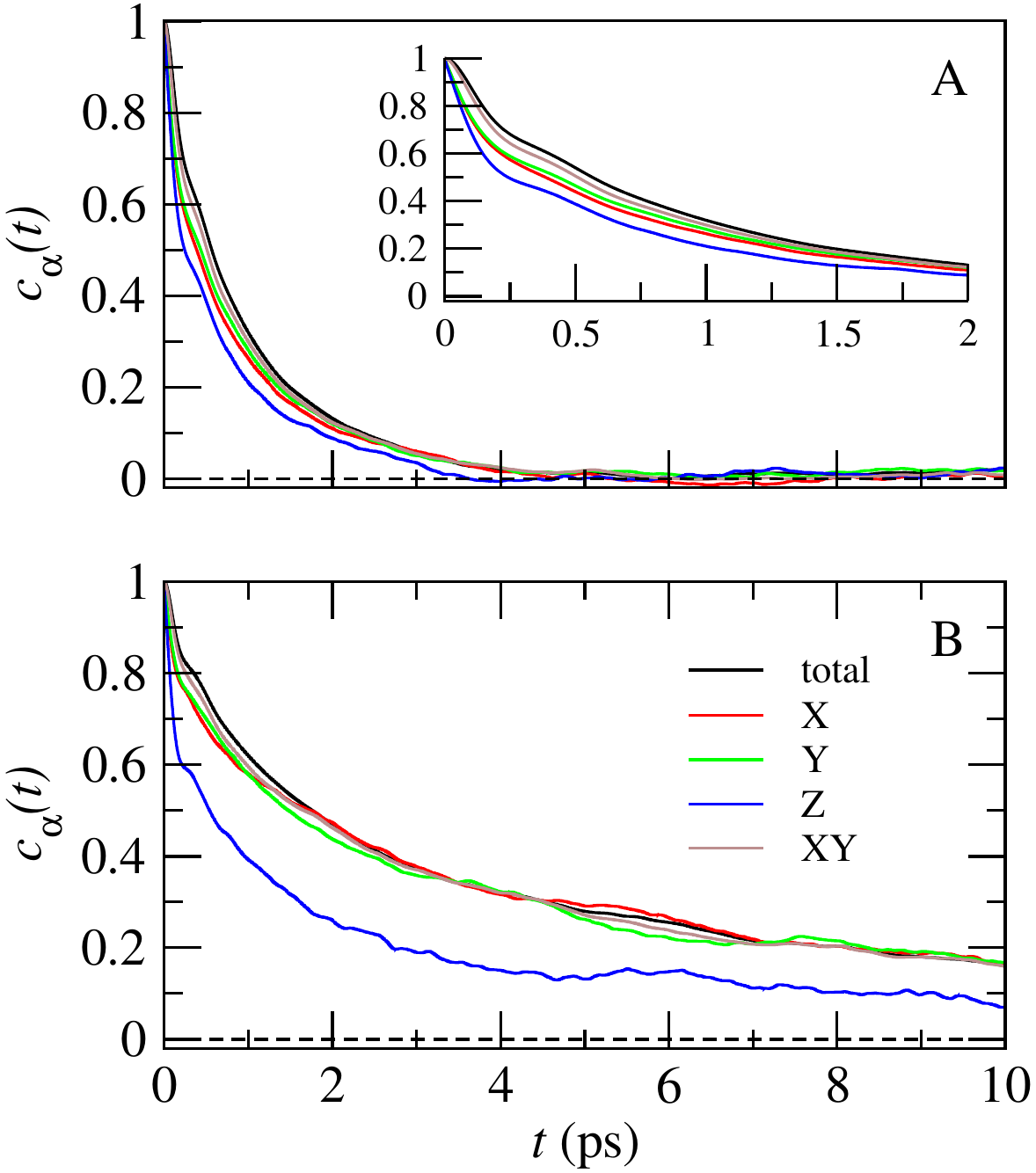}
\caption{Self correlation functions for the C-N bond vectors of the MA molecules calculated from A) System 333 and B) System 222 simulated trajectories. The inset shows a detail at short times for System 333.}
\label{corres}
\end{figure}

In order to understand the dynamical behavior of the MA molecules we calculated the time dependent self correlation function of the C-N bonds. Namely, we calculated the function $c(t)=\langle \vec b(t) \cdot \vec b(0) \rangle$ with $\vec b$ being a unit vector pointing from the C atom toward the N atom of the same MA molecule. The angular brackets represent the time average. We also calculated $c_x(t)$, $c_y(t)$, $c_z(t)$ and $c_{xy}(t)$ using the components of the vector $\vec b$ indicated by the subscripts. The results are presented in Figure \ref{corres} for the two systems. There is important difference between the two sets of results pointing to finite size effects affecting the dynamics of the MA molecules. While for the System 333 the relaxation times are $\sim$ 1.5 ps, for System 222 the curves do not reach a totally  decorrelated state for the investigated times, i.e. the curves for System 222 do not reach zero after 10 ps. Both sets of results where obtained using the final 35 ps of simulations, so the data extends up to 17.5 ps, yet the final 7.5 ps tail of the correlation curves do not add significant information and was dropped from the plots. The question that arises is what is the origin of the difference between the two results. In order to answer this we look at the relaxation of the individual components of $\vec b$ in the two systems. In the larger system all the $c_\alpha (t)$ curves are very similar to each other, but in the smaller system  $c_z (t)$ relaxes much faster than the other curves. It could be argued that the origin of the stronger correlation in the smaller system is a consequence of the effects of the electrostatic interactions. In fact, the long range electrostatic interactions may artificially increase the correlations between neighboring MA molecules resulting in a slower rotational mode of each individual MA molecule. This argument is supported by the findings from System 222 and the tetragonal shape of the simulation box. The average dimensions of the box in $a_x=a_y=20.0$ \AA\ and $a_z=25.7$ \AA\ and then the electrostatic interaction have more space to relax along the $z$ direction. On the contrary, the shorter dimensions along $x$ and $y$ imposes a periodicity that when contrasted with the results of the larger system appears as unrealistic. For reference, the average dimensions of the simulation box for System 333, calculated from the last 20 ps of simulations are $a_x=a_y=27.2$ \AA\ and $a_z=38.9$ \AA.

\begin{figure}[!t]
\includegraphics*[width=0.45\textwidth]{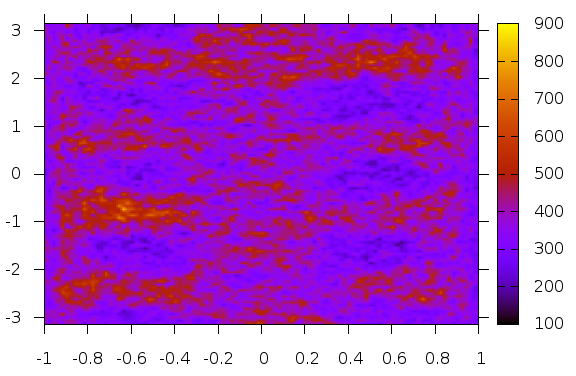}
\caption{Density map of the orientation of the C-N bonds of the MA molecules. The horizontal and vertical axis represent $\cos \theta$ and $\phi$, respectively. The color scale is proportional to the average time spend by the molecules at a given orientation.}
\label{densmap}
\end{figure}

An important feature that is also revealed by the correlation curves is a small kink at $t \simeq 0.35$ ps that is clearly visible in the inset of Figure \ref{corres}. This kink, although small, is present in all the curves represents another indication of anisotropic rotations of the MA molecules. We further explored the rotational pattern of the MA molecules by calculating the density map of the direction $(\cos \theta,\phi)$ of the vectors $\vec b$. Here we define $\theta$ and $\phi$ as the spherical coordinates of $\vec b$ in the frame defined by the simulation box. Discarding the first 5 ps of simulation we determined the directional angles at each time step for all the MA molecules and then we calculated the $(\cos \theta,\phi)$ density map on the $[-1:1,-\pi:\pi]$ plane that is displayed in Figure \ref{densmap}. The anisotropy in the orientation of the MA molecules, although weak, is reflected by four denser stripes for $|\cos \theta| > 0.3$. This, plus the fact that the stripes are four and regularly distributed along the $\phi$ domain suggest that the MA molecules have a (slightly) higher probability of pointing diagonally in their inorganic cages. Moreover, the diagonal configuration allows for the molecules to gain conformational entropy since they have more space to rotate along the C-N bond and then it is expected that the MA molecules spend more time on those orientations. The corresponding density map from System 222, included in the Supporting Material, shows a heavier weight at the bottom of the plot than at the top, in line with the artificially longer correlation times.

\begin{figure}[!t]
\includegraphics*[width=0.34\textwidth]{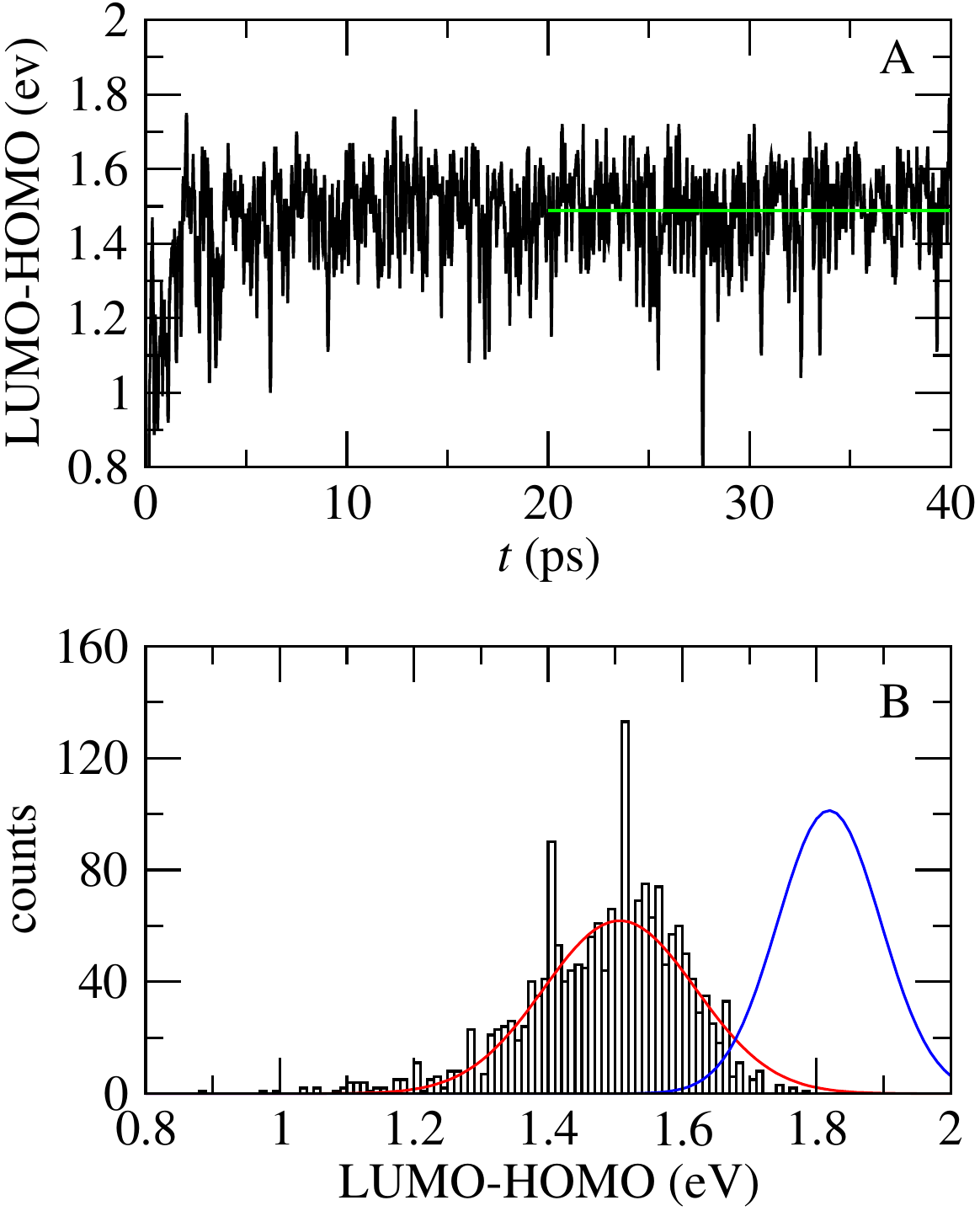}
\caption{A) LUMO-HOMO energy difference as a function of time for System 333. The green line represents the average of the last 20 ps of simulation at 1.489 eV. B) Histogram of the energy gap for the last 35 ps of simulation. The red line is the best fit using a normal distribution with $\mu=1.506$ eV and $\sigma=0.113$ eV. The blue line is the corresponding result for System 222.}
\label{gap}
\end{figure}

\begin{figure}[!t]
\includegraphics*[width=0.4\textwidth]{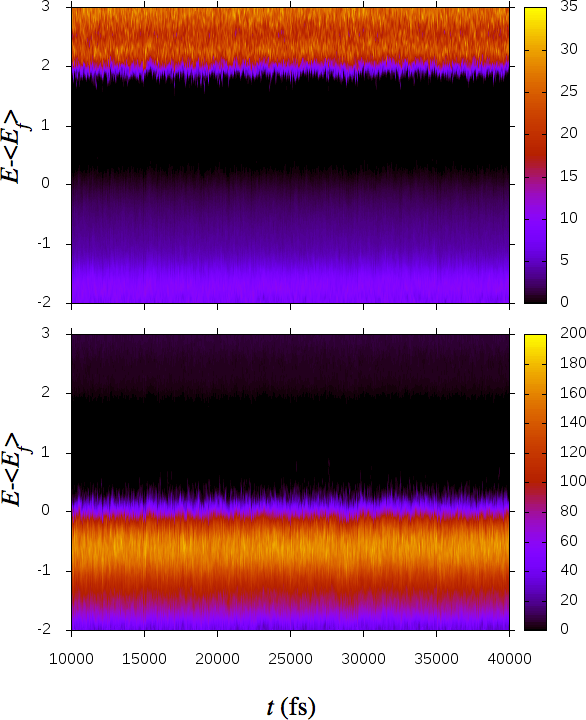}
\caption{The bottom panel shows the PDOS for the $p_z$ of I atoms (bottom) and Pb (top). The horizontal axis corresponds to the simulation time, the vertical axis is $E-\langle E_f \rangle$, with $\langle E_f \rangle$ calculated as the time average (10 to 40 ps) of the energy of the HOMO. The PDOS is represented by the color scale.}
\label{pros}
\end{figure}

We now turn our attention to the electronic structure properties of the MAPbI$_3$ perovskite that is the most studied subject from a theoretical point of view for these materials. The key property that determine the optical absorption is the band gap energy, experimentally determined to be 1.55 eV \cite{Baikie:2013aa}. In Figure \ref{gap} we show the LUMO-HOMO energy difference for System 333 as a function of time, along with a histogram showing the distribution of energies. The calculated energy gaps fluctuate around $\mu=1.506$ eV following approximately a normal distribution with standard deviation $\sigma=0.113$ eV. For reference we include the corresponding fit for System 222, which is centred at higher energies. The important difference between the two results is a consequence of the $\Gamma$-point approximation that strongly depends on the system size. Further information can be obtained monitoring the projected density of states (PDOS) as a function of time, which was computed every 20 fs along the whole simulation trajectory. Visual inspection of the results points out that at the Fermi level the $p$ orbitals corresponding to the I atoms have the higher density. Actually, the three components ($p_x$, $p_y$ and $p_z$) of the I atoms are very similar to each other and for that reason we plot only one of them in Figure \ref{pros}. The lowest virtual orbitals with higher PDOS correspond to the $p$ orbitals of the Pb atoms. Similarly to the I case, the $p_i$ components are all similar and we plot only one on Figure \ref{pros}. Our findings are in line with previous works based on static DFT calculations and show that the exciton creation involves the transition of an electron from an I atom to a Pb atom. It is interesting to note that, although the LUMO-HOMO energy gap fluctuates considerably around its central value, the edges of the PDOS shown on Figure \ref{pros} are smooth. Inspecting the numerical results we find that one source of variations in the calculated gap is that in some cases the smallest gap occurs involving atoms far apart in the system; namely, involving atoms that are not first neighbors of each other. The creation of an electron-hole pair  leaving the hole very far apart from the exited electron is unlikely and therefore the low energy tail of the histogram showed in Figure \ref{gap} is perhaps artificial.

\section{Conclusions}

We have performed a molecular dynamic simulation study on the prototypical example of organic-inorganic perovskite used for experimental solar cells, i.e. CH$_3$NH$_3$PbI$_3$. To the best of our knowledge, this study represent the first effort addressing the tetragonal structure that is the stable structure at ambient standard conditions.  The comparison of the local structure averaged along the simulation with the experimentally determined crystal structure demonstrate the stability of the system giving confidence in the overall computational approach. In spite of an expansion of the simulated system of about 2.6 \% with respect to the experimental volume, the local structure of the inorganic scaffold is maintained without modifications. The methylammonium molecules, instead, rotate within their confining cages showing a slight anisotropic behavior. 

An important part of this work is the analysis of the finite size effects affecting the dynamics of the MA molecules. We clearly show that a system involving a 2$\times$2$\times$2 replication of the unit cell is insufficient to allow for the proper decorrelation between neighboring molecules. This conclusion is very important and clearly affects the conclusion of published works involving similar systems to the one that we have studied. We believe that the implications of this finding goes beyond perovskite systems in particular and should apply to molecular ionic crystals and ionic liquids where the main attractive forces result from the charge separations and are affected by long range correlations. 

\section{acknowledgments}

The HPC resources and services used in this work were provided by the Research Computing group in Texas A\&M University at Qatar. Research Computing is funded by the Qatar Foundation for Education, Science and Community Development.

\newpage



\section*{Supplementary material}

# Thermal effects on CH3NH3PbI3 perovskite from *ab-initio* molecular dynamics

# Supplementary Material

Marcelo A. Carignano and Ali Kachmar
Qatar Environment and Energy Research Institute, Qatar Foundation, P.O. Box 5825, Doha, Qatar

and

Jürg Hutter
Department of Chemistry, University of Zurich, Winterthurerstrasse 190, CH-8057, Zurich, Switzerland

### *SNAPSHOTS*

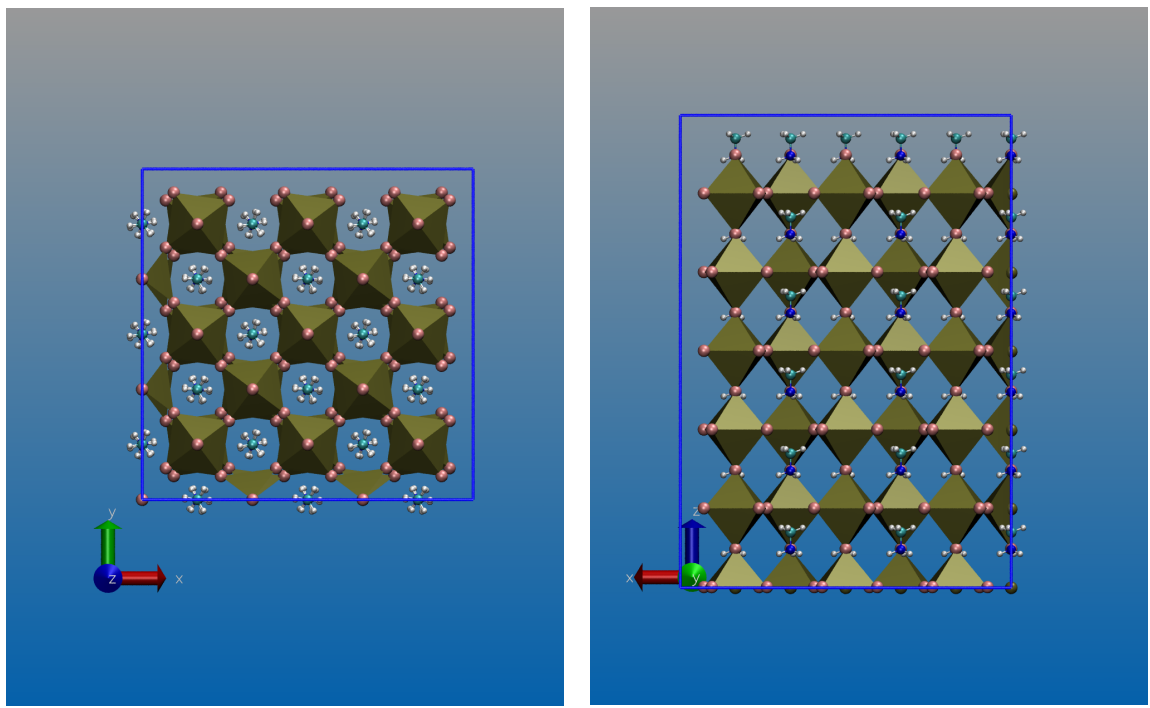

Figure S1. Snapshots of the initial structure of the simulations for System 333 obtained from the CIF files published by Stoumpos et al., *Inorganic Chemistry* **52**, 9019 (2013).

## PAIR DISTRIBUTION FUNCTIONS

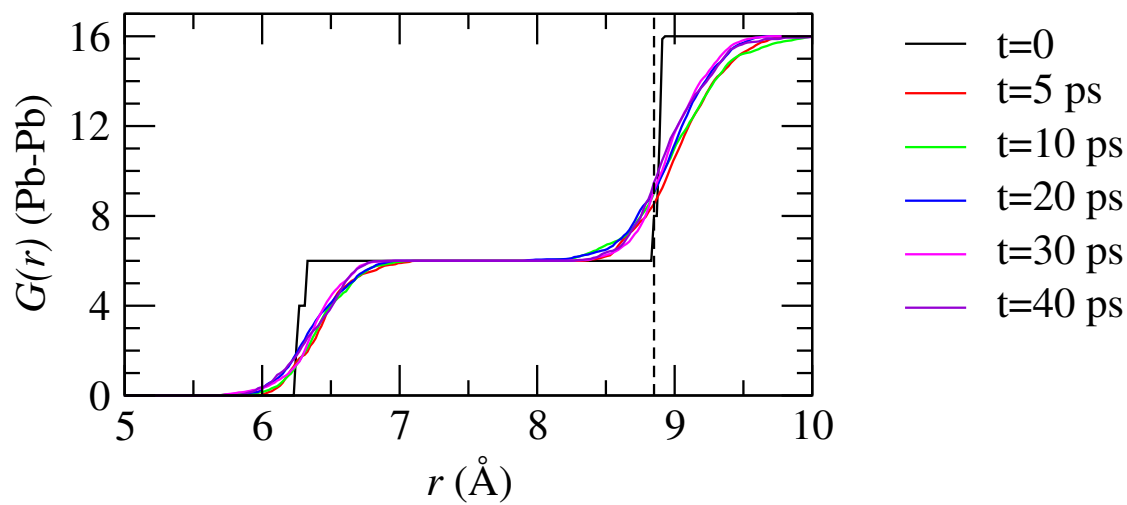

Figure S2. Cumulative pair distribution function for Pb-Pb atoms calculated from the simulation of System 222. The different colors correspond to different times along the simulation as indicates. The vertical dashed line represent the maximum value for $r$ allowed by the minimal image convention. Notice that this system measure incorrectly the number of second neighbors, which should be 12. Then G(r~10)=18 and 16. Compare with the results from Figure 4 of the main paper.

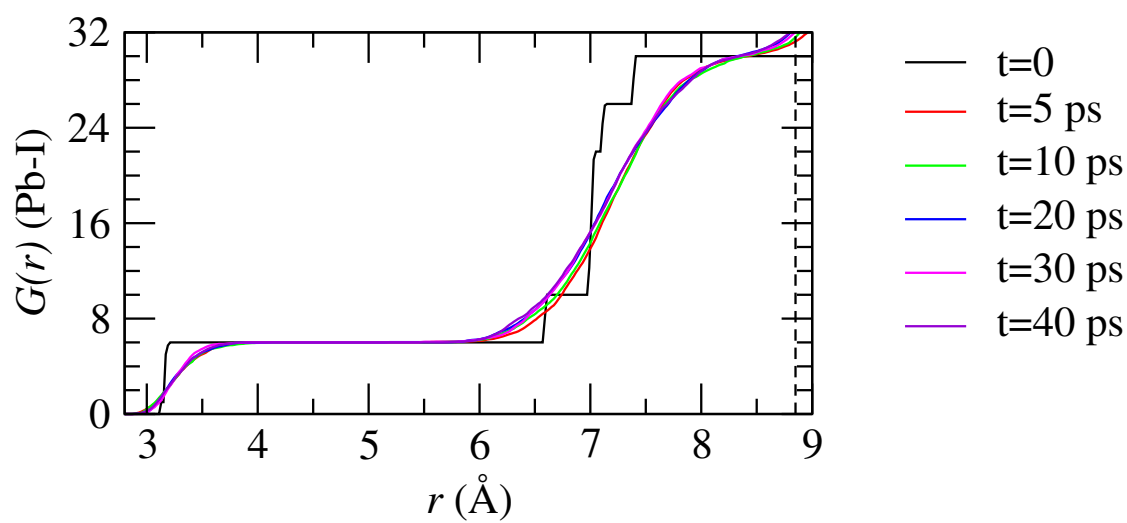

Figure S3. Cumulative pair distribution function for Pb-I atoms calculated from the simulation of System 222.

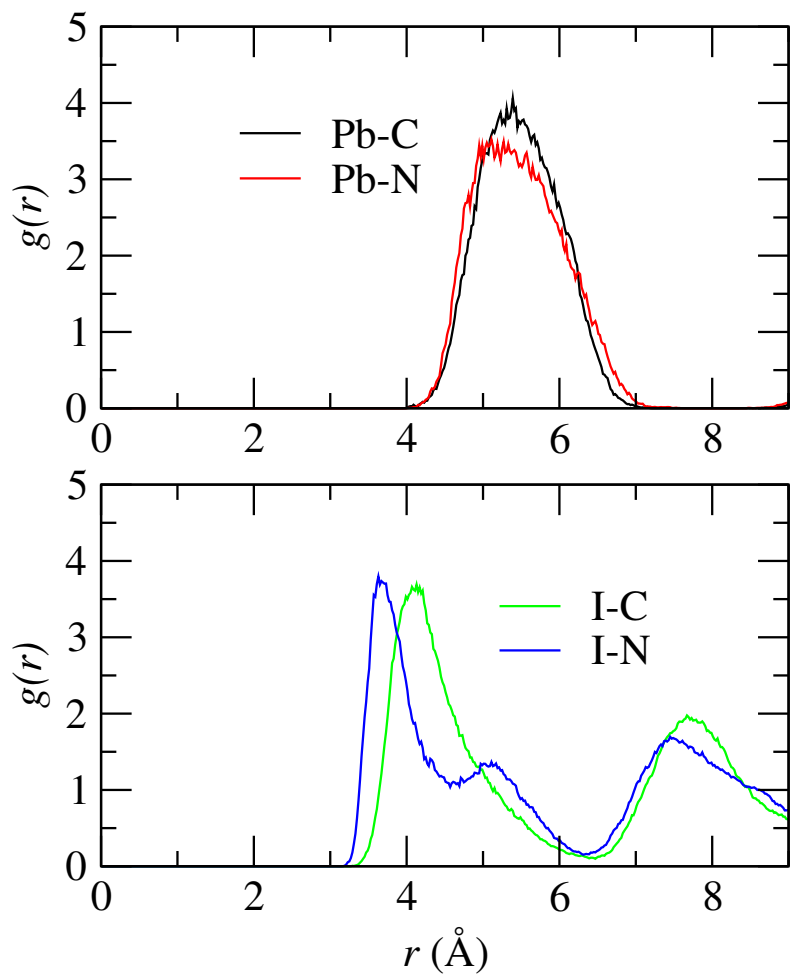

Figure S4. Pair distribution functions for Pb-C, Pb-N (top panel) and I-C and I-N (bottom panel) calculated from the simulation of System 222. The curves were calculated using the last 5 ps of simulation.

## ORIENTATION OF MA MOLECULES

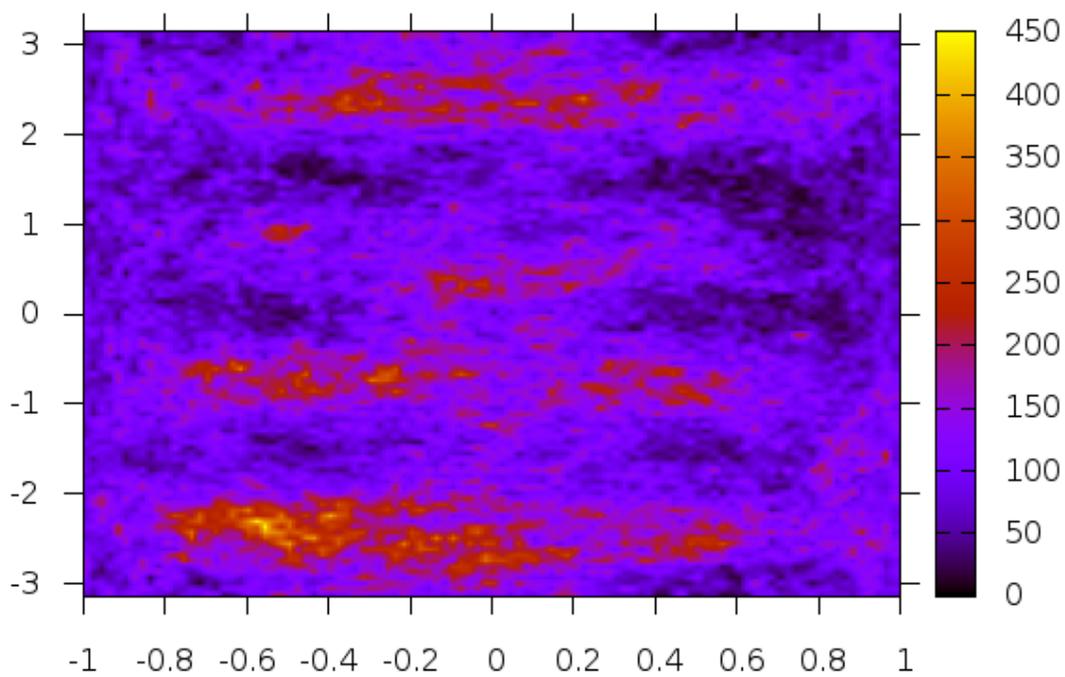

Figure S5. Density map obtained from the orientation (cos $\theta$, $\varphi$) calculated from the simulation of System 222. The slower rotational dynamics of this system is reflected by the difference between the top half (positive $\varphi$) and bottom half of the figure.

*TEST OF ANISOTROPIC NPT ENSEMBLE*

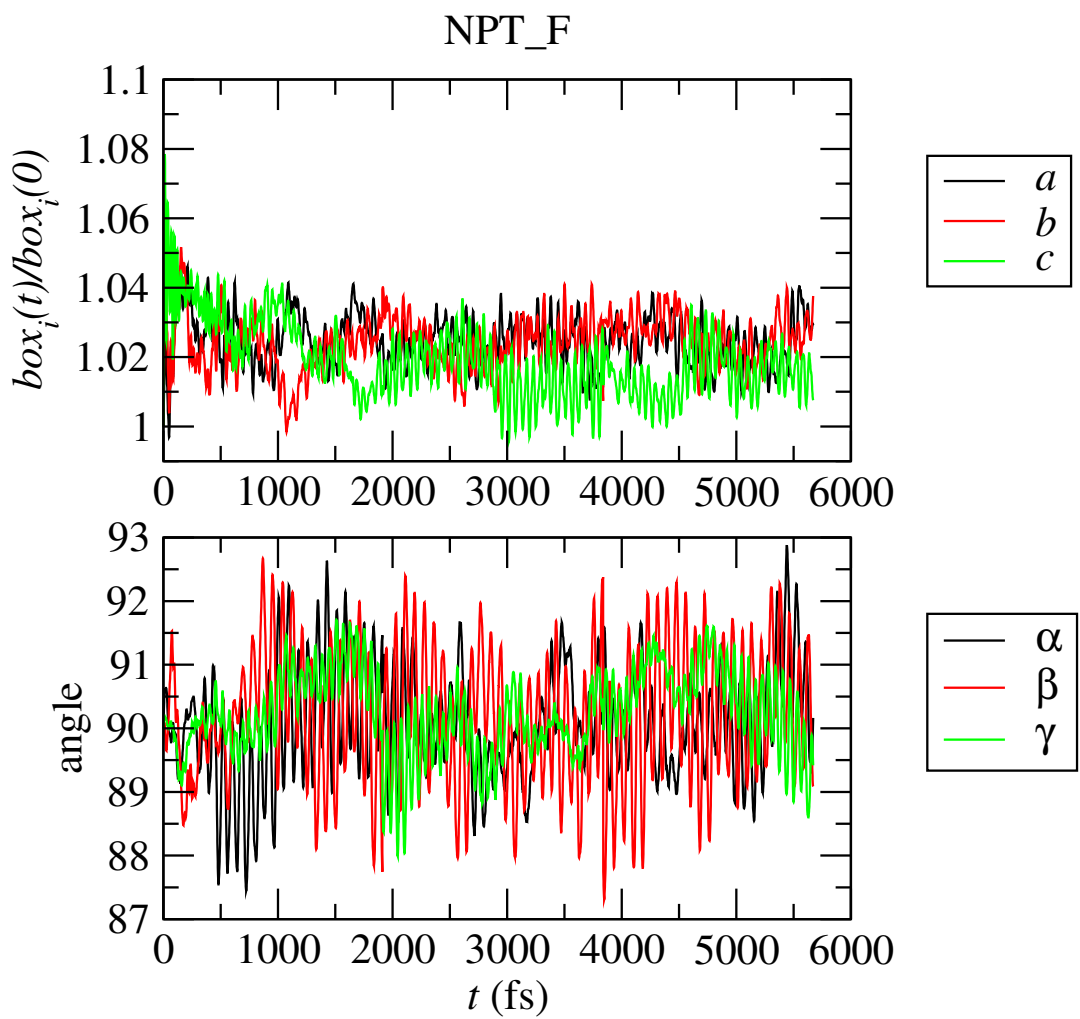

Figure S6. Relative box dimension and angles from a test simulation using the NPT_F ensemble option of CP2K. The three box dimensions have a similar expansion factor, and the three angles fluctuate around 90, ensuring that the isotropic constraint is adequate.

## PROJECTED DENSITY OF STATES

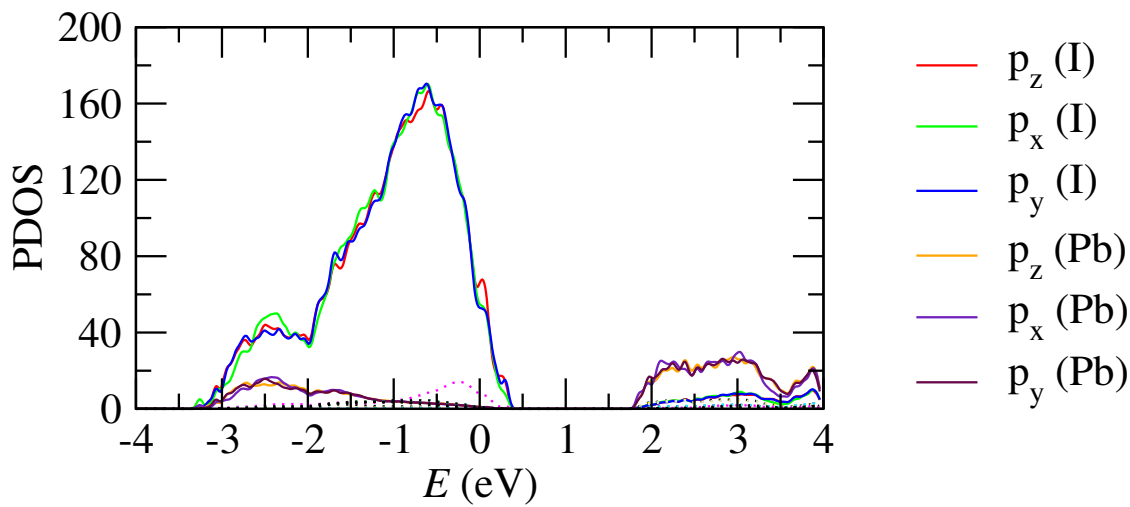

Figure S7. Projected density of states (PDOS) calculated at t=25 ps from the simulation of System 333. The figure highlight the three p orbitals of the I and Pb, which are the largest at the edges of the energy gap.

## HYDROGEN BONDS

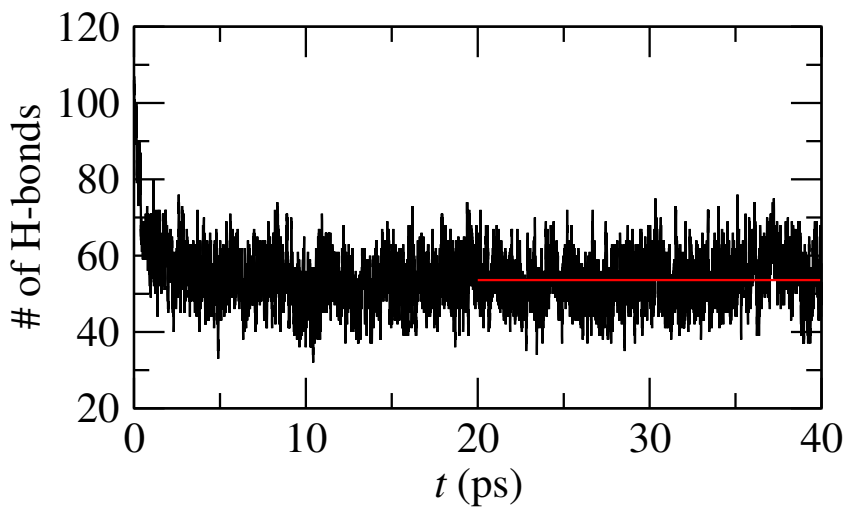

Figure S8. Number of H bonds between N and I atoms for System 333 along the whole trajectory. The red line shows the average number of bonds (53.6) in the second 20 ps of simulation. H-bonds were defined for distance(I-N) < 4.6 Å and angle(I-N-A) < 20°. The distance cutoff corresponds to the first minimum of the I-N pair distribution function, Figure 6 of the main paper.

## *ENERGY GAP*

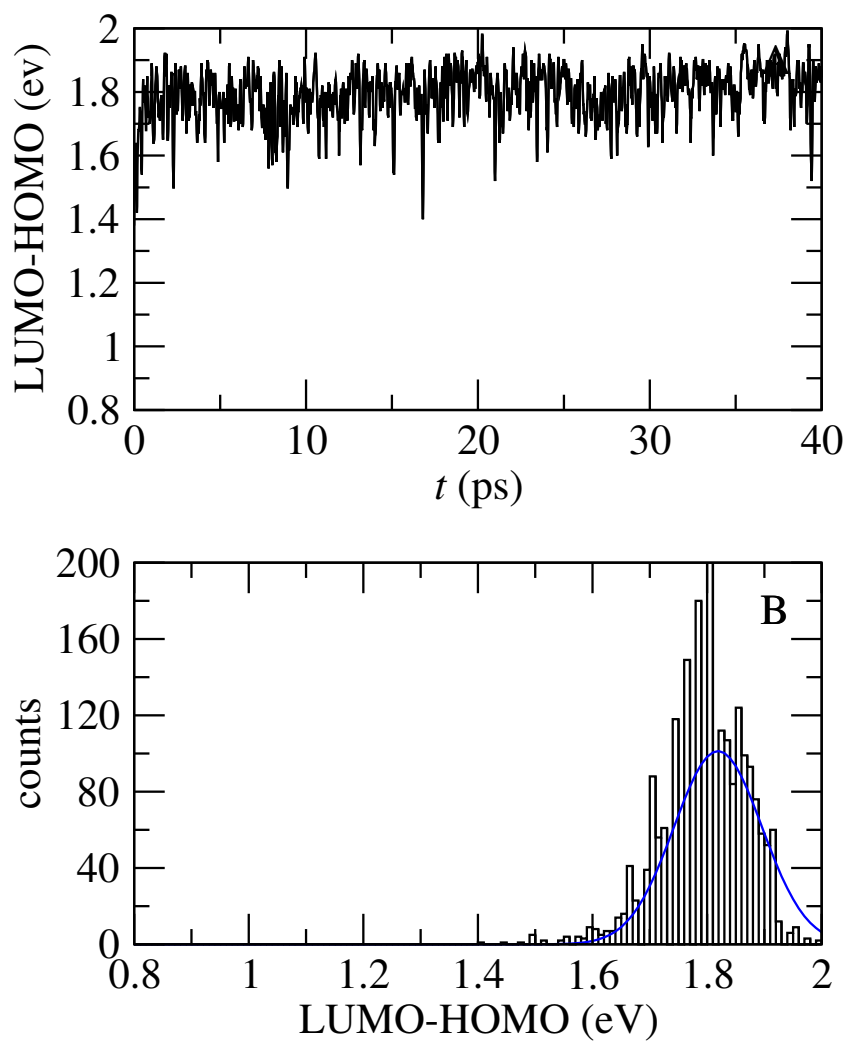

Figure S9. A) LUMO-HOMO energy difference as a function of time for System 222. B) Histogram and fitting with a normal curve (blue line), with parameters $\mu$=1.81 eV and $\sigma$=0.07 eV.



\end{document}